\documentclass[twocolumn,trackchanges]{aastex63}

\usepackage{amsmath}

\usepackage{verbatim}

\shortauthors{De Graaff et al.}
\graphicspath{{./}{figures/}}
\begin{document}

\title{Tightly coupled morpho-kinematic evolution for massive star-forming and quiescent galaxies across 7 Gyr of cosmic time}

\author{Anna de Graaff}\affiliation{Leiden Observatory, Leiden University, P.O.Box 9513, NL-2300 AA Leiden, The Netherlands; \url{ graaff@strw.leidenuniv.nl}}

\author{Rachel Bezanson}\affiliation{Department of Physics and Astronomy, University of Pittsburgh, Pittsburgh, PA 15260, USA}

\author{Marijn Franx}\affiliation{Leiden Observatory, Leiden University, P.O.Box 9513, NL-2300 AA Leiden, The Netherlands; \url{ graaff@strw.leidenuniv.nl}}

\author{Arjen van der Wel}\affiliation{Sterrenkundig Observatorium, Universiteit Gent, Krijgslaan 281 S9, B-9000 Gent, Belgium}\affiliation{Max-Planck-Institut f\"ur Astronomie, K\"onigstuhl 17, D-69117, Heidelberg, Germany}

\author{Eric F. Bell}\affiliation{Department of Astronomy, University of Michigan, 1085 S. University Avenue, Ann Arbor, MI 48109, USA}

\author{Francesco D'Eugenio}\affiliation{Sterrenkundig Observatorium, Universiteit Gent, Krijgslaan 281 S9, B-9000 Gent, Belgium}

\author{Bradford Holden}\affiliation{UCO/Lick Observatory, University of California, Santa Cruz, CA 95064, USA}

\author{Michael V. Maseda}\affiliation{Leiden Observatory, Leiden University, P.O.Box 9513, NL-2300 AA Leiden, The Netherlands; \url{ graaff@strw.leidenuniv.nl}}

\author{Adam Muzzin}\affiliation{Department of Physics and Astronomy, York University, 4700 Keele St., Toronto, Ontario, M3J 1P3, Canada}

\author{Camilla Pacifici}\affiliation{Space Telescope Science Institute, 3700 San Martin Drive, Baltimore, MD 21218, USA}

\author{Jesse van de Sande}\affiliation{Sydney Institute for Astronomy, School of Physics, A28, The University of Sydney, NSW, 2006, Australia}\affiliation{ARC Centre of Excellence for All Sky Astrophysics in 3 Dimensions (ASTRO 3D), Australia}

\author{David Sobral}\affiliation{Department of Physics, Lancaster University, Lancaster LA1 4YB, UK}

\author{Caroline M.S. Straatman}\affiliation{Sterrenkundig Observatorium, Universiteit Gent, Krijgslaan 281 S9, B-9000 Gent, Belgium}

\author{Po-Feng Wu}\affiliation{National Astronomical Observatory of Japan, 2-21-1 Osawa, Mitaka, Tokyo 181-8588, Japan}



% max 250 words!
\begin{abstract}

We use the Fundamental Plane (FP) to measure the redshift evolution of the dynamical mass-to-light ratio ($M_{\mathrm{dyn}}/L$) and the dynamical-to-stellar mass ratio ($M_{\mathrm{dyn}}/M_*$). Although conventionally used to study the properties of early-type galaxies, we here obtain stellar kinematic measurements from the Large Early Galaxy Astrophysics Census (LEGA-C) Survey for a sample of $\sim1400$ massive ($\log( M_*/M_\odot) >10.5$) galaxies at $0.6<z<1.0$ that span a wide range in star formation activity.
In line with previous studies, we find a strong evolution in $M_{\mathrm{dyn}}/L_g$ with redshift. In contrast, we find only a weak dependence of the mean value of $M_{\mathrm{dyn}}/M_*$ on the specific star formation rate, and a redshift evolution that likely is explained by systematics. Therefore, we demonstrate that star-forming and quiescent galaxies lie on the same, stable mass FP across $0<z<1$, and that the decrease in $M_{\mathrm{dyn}}/L_g$ toward high redshift can be attributed entirely to evolution of the stellar populations. Moreover, we show that the growth of galaxies in size and mass is constrained to occur within the mass FP. Our results imply either minimal structural evolution in massive galaxies since $z\sim1$, or a tight coupling in the evolution of their morphological and dynamical properties, and establish the mass FP as a tool for studying galaxy evolution with low impact from progenitor bias.

\end{abstract}



\section{Introduction} \label{sec:intro}

Galaxies obey a tight scaling relation between size, velocity dispersion, and surface brightness or stellar mass surface density, known as the Fundamental Plane (FP) 
\citep[e.g.,][]{Djorgovski1987,Dressler1987,Jorgensen1996}. The tilt and zero point of the luminosity FP are directly related to the dynamical mass-to-light ratio ($M_{\mathrm{dyn}}/L$) \citep{Faber1987}, and the FP has therefore proven to be a valuable tool in studying the evolution in $M_{\mathrm{dyn}}/L$ of the quiescent galaxy population. The zero point in particular has been shown to evolve significantly with redshift, which 
places strong constraints on the formation epoch of massive quiescent galaxies \citep[e.g.,][]{vanDokkum1996,vdWel2005}.

However, \citet{Saglia2010,Saglia2016} and \citet{Toft2012} suggest that evolution in the morphological or kinematic structure may be required to fully account for the observed evolution in the FP. \citet{Bezanson2013}, on the other hand, demonstrate that when the surface brightness parameter in the FP is replaced by the stellar mass surface density, there is very little evolution in the resulting mass FP of massive quiescent galaxies to $z\sim2$. These observations suggest that any redshift dependence of $M_{\mathrm{dyn}}/L$ is caused primarily by evolution in the stellar mass-to-light ratio ($M_*/L$), %with minimal change in the structure-dependent ratio of the total and stellar mass ($M_{\mathrm{dyn}}/M_*$).  
and that changes in the structure-dependent ratio of the total and stellar mass ($M_{\mathrm{dyn}}/M_*$) are either minimal, or embedded in the FP. 

High-redshift studies of the FP thus far, however, have been limited in sample size, with selections being biased toward either the densest environments or brightest objects \citep[e.g.,][]{Holden2010,vdSande2014,Beifiori2017, Prichard2017,Saracco2020}, which populate the FP differently than typical galaxies in the field \citep[see, e.g.,][]{Saglia2010,vdSande2014}. Extending these analyses to a more representative sample of the overall galaxy population is therefore crucial to understand the redshift evolution in $M_{\mathrm{dyn}}/L$ and $M_{\mathrm{dyn}}/M_*$.

At low redshift, \citet{Zaritsky2008} and \citet{Bezanson2015} have shown that star-forming galaxies lie on the same surface as the quiescent galaxies, if both $M_*/L$ and rotation velocities are taken into account. In de Graaff et al. (in prep.) we present the luminosity and mass FP of a large, $K_{\mathrm{s}}$-band selected sample of galaxies drawn from the Large Early Galaxy Astrophysics Census (LEGA-C) Survey \citep{vdWel2016,Straatman2018}, and find that star-forming and quiescent galaxies also lie on the same mass FP at $z\sim0.8$. %, despite significant differences in their $M_{\mathrm{dyn}}/L$ ratios.

In this Letter, we constrain the redshift evolution of the luminosity FP and mass FP between $0<z<1$, by using our representative sample of massive galaxies from the LEGA-C survey and a reference sample of local galaxies from the Sloan Digital Sky Survey (SDSS). 

We assume a flat $\Lambda$CDM cosmology throughout, with $\Omega_{\mathrm{m}}=0.3$ and $H_0=70\,\mathrm{km\,s^{-1}\,Mpc^{-1}}$. %All magnitudes are in the AB photometric system.

 \section{Data} \label{sec:data}

\subsection{LEGA-C Survey}\label{sec:legac}

Our sample is drawn from the third data release of the LEGA-C survey, a deep spectroscopic survey of $\sim3000$ $K_{\mathrm{s}}$-band selected galaxies at $0.6<z<1.0$ in the COSMOS field \citep{vdWel2016,Straatman2018}, which provides accurate absorption line widths for a representative sample of the massive galaxy population at $z\sim0.8$. 

We describe the combined data set and our sample selection in detail in de Graaff et al. (in prep.). Briefly, we measure integrated stellar velocity dispersions, to which both the intrinsic velocity dispersion and projected rotational motions contribute, from the LEGA-C spectra \citep[see][]{Bezanson2018b, Straatman2018}. We obtain structural parameters by fitting S\'ersic profiles to ACS F814W imaging from the \textit{Hubble Space Telescope} with GALFIT \citep{Peng2010}, and circularize the effective radii (i.e. $R_\mathrm{e}=\sqrt{ab}$). We derive stellar masses by fitting the galaxy spectral energy distributions (SEDs) with MAGPHYS \citep{daCunha2008} and measure rest-frame luminosities with EAZY \citep{Brammer2008}, using the multi-wavelength ($0.2-24\,\micron$) photometric catalog by \citet{Muzzin2013a}. We correct all masses and luminosities for missing flux using the total luminosity of the best-fit S\'ersic profile \citep[e.g.,][]{Taylor2010}.

We select galaxies of stellar mass $\log(M_*/M_\odot)\geq10.5$, and require a maximum uncertainty of $15\%$ on the velocity dispersion. Moreover, we require that the GALFIT fit has converged, and remove galaxies which are significantly morphologically disturbed. Our final sample consists of 1419 galaxies. We use the rest-frame $U-V$ and $V-J$ colors and the selection criteria by \citet{Muzzin2013b} to define quiescent and star-forming subsamples.

\subsection{SDSS}\label{sec:sdss}

We obtain a reference sample of galaxies at $0.05<z<0.07$ from the 7\textsuperscript{th} data release of the SDSS \citep{sdss:dr7}, matching the selection criteria and observables as closely as possible to the LEGA-C sample. Our selection and aperture corrections are detailed in de Graaff et al. (in prep.). Briefly, we require a maximum uncertainty on the stellar velocity dispersion of $15\%$, and correct the velocity dispersions to an aperture of $1\,R_\mathrm{e}$. We use stellar masses estimated from SED fitting with MAGPHYS \citep{Chang2015}, and structural parameters derived from single S\'ersic models in the $r$-band \citep{Simard2011}. We circularize the effective radii, and correct all stellar masses using the total luminosity of the best-fit S\'ersic profile. Our selection consists of 23,036 massive galaxies ($\log(M_*/M_\odot)\geq10.5$).

\begin{figure*}
    \centering
    \includegraphics[width=0.9\linewidth]{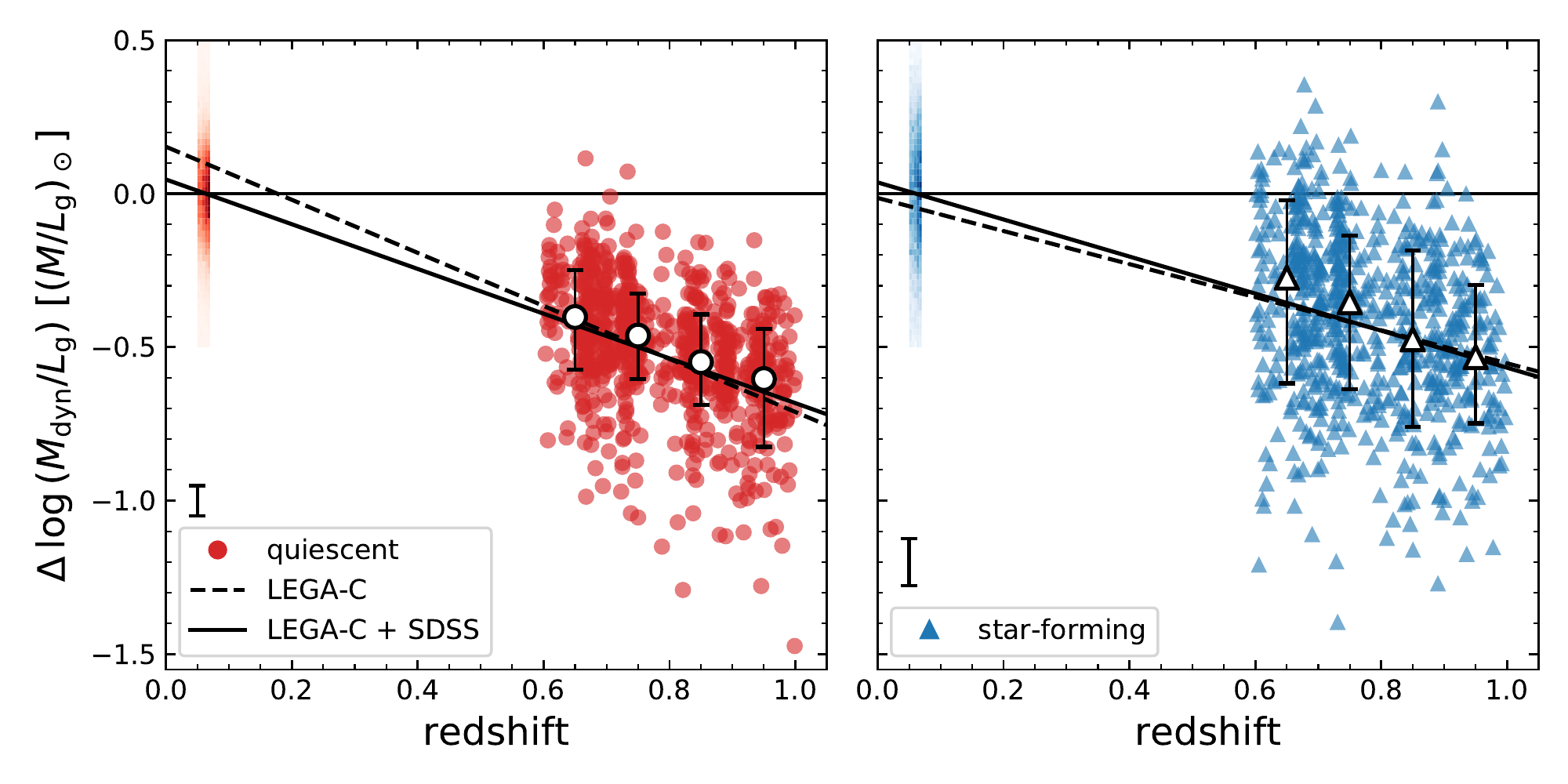}
    \caption{Redshift evolution of the dynamical mass-to-light ratio of quiescent (left) and star-forming (right) galaxies from the SDSS and LEGA-C samples. Linear fits to the LEGA-C data alone (dashed lines) and combined LEGA-C and SDSS sample (solid lines) show that there is a strong evolution in $M_{\mathrm{dyn}}/L_g$ with redshift, with the quiescent population evolving more rapidly than the star-forming population (Table~\ref{tab:zevo}). }
    \label{fig:ML_evo}
\end{figure*}

Rest-frame colors and luminosities are calculated using \textsc{kcorrect} \citep{Blanton2007}, and we differentiate between quiescent and star-forming galaxies using the rest-frame $u-r$ and $r-z$ colors and the criteria from \citet{Holden2012}.

\section{Evolution in $M_{\mathrm{dyn}}/L$} \label{sec:ML_results}

The FP in luminosity, here taken as the rest-frame $g$-band luminosity, has the form: 
\begin{equation}
    \log R_\mathrm{e}=a\,\log\sigma+b\,\log I_{\mathrm{ e,g}}+c,
     \label{eq:lfp}
\end{equation}
where $R_\mathrm{e}$ is the effective radius, $\sigma$ the integrated stellar velocity dispersion, and $I_\mathrm{e,g}=-0.4\,\mu_\mathrm{e,g}$, where $\mu_\mathrm{e,g}$ is the mean surface brightness within the effective radius, corrected for cosmological surface brightness dimming \citep[see, e.g.,][]{HydeBenardi2009}. The coefficients $a$ and $b$ describe the tilt of the plane, and $c$ is the zero point.

We assume that the tilt of the FP does not evolve strongly with redshift \citep[as shown in][de Graaff et al. in prep.]{Holden2010}, and adopt the tilt derived by \citet{HydeBenardi2009}, of $a=1.404$ and $b=-0.761$, for both the SDSS and LEGA-C samples.  
We fit the zero point $c$ of the FP for the SDSS sample by minimizing the mean absolute orthogonal deviations from the FP,
\begin{equation}
    \Delta_{\mathrm{FP}}=\frac{|\log\,R_\mathrm{e}-a\,\log\,\sigma-b\,\log\,I_{\mathrm{e,g}}-c\,|}{\sqrt{1+a^2+b^2}}.
    \label{eq:deltaFP}
\end{equation} 
Next, we determine for each LEGA-C galaxy the difference in $\log(M_{\mathrm{dyn}}/L_g)$ with respect to the SDSS sample, by firstly calculating the residual of the FP in $\log I_{\mathrm{e,g}}$:
\begin{equation}
    \Delta\log I_{\mathrm{e,g}}=-\left(\Delta_{\mathrm{LFP}}-c_0\right)/b,
    \label{eq:ML}
\end{equation}
where $c_0$ is the best-fit zero point to the SDSS data, and
\begin{equation}
    \Delta_{\mathrm{LFP}}=\log R_\mathrm{e}-a\,\log\sigma-b\,\log I_{\mathrm{e,g}}.
\end{equation}
We then make the common assumption that $\Delta\log I_{\mathrm{e,g}}$ is dominated by variations in $M_{\mathrm{dyn}}/L$:
\begin{equation}
    \Delta\log(M_{\mathrm{dyn}}/L_\mathrm{g})\approx-\Delta\log I_{\mathrm{e,g}}.
\end{equation}

We perform these calculations separately for the quiescent and star-forming populations, and show the observed redshift evolution of $M_{\mathrm{dyn}}/L_g$ in Fig.~\ref{fig:ML_evo}. Similar to many previous FP studies of quiescent galaxies \citep[e.g.,][]{vdWel2005,VanDokkum2007}, we find that $ M_{\mathrm{dyn}}/L_g$ decreases with redshift, and show that this is also the case for the star-forming population. We determine the slope of the redshift evolution using a linear least squares fit, weighted by the observational errors, and estimate uncertainties on the fit via bootstrap resampling. The number of SDSS galaxies is significantly larger than the LEGA-C sample size, which effectively causes the fit to be forced through the best-fit zero point of the SDSS FP ($\Delta\log(M_{\mathrm{dyn}}/L_g)=0$). Since this omits any potential systematic errors on the SDSS data, we fit to both the combined LEGA-C and SDSS data (solid lines) and the LEGA-C data only (dashed lines).

The results are presented in Table~\ref{tab:zevo}; the two different methods agree within $2\sigma$ and $1\sigma$ for the quiescent and star-forming samples respectively. Some small systematic discrepancies between the two different estimates for each subsample may be expected, considering that there are substantial differences in the measurements of the effective radii, velocity dispersions, and photometry between the SDSS and LEGA-C data.

\subsection{Quiescent galaxies}\label{sec:zevo_LFP_Q}

\begin{figure}
    \centering
    \includegraphics[width=\linewidth]{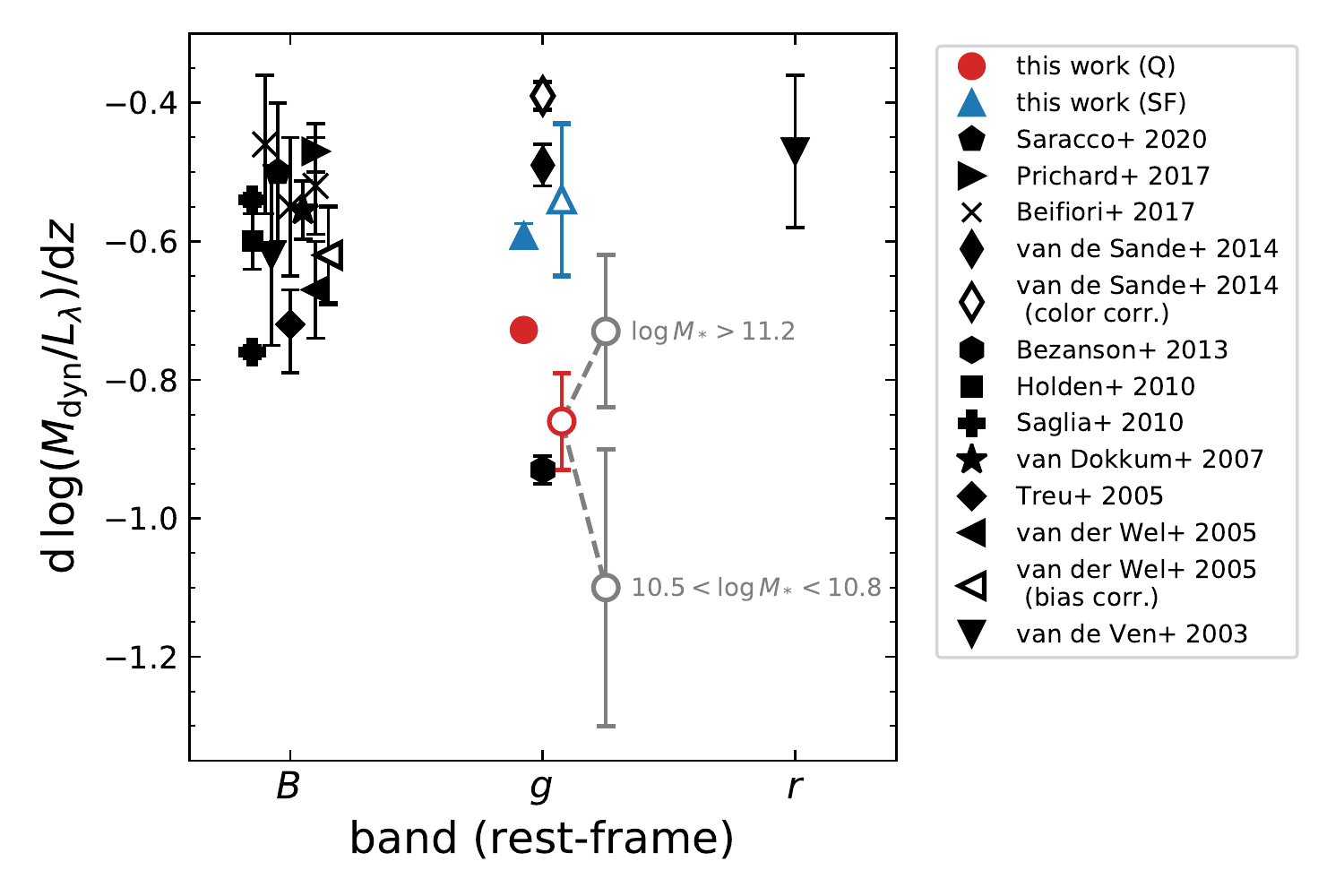}
    \caption{Comparison of the measured redshift evolution in $M_{\mathrm{dyn}}/L$ in different passbands. Red and blue markers show the results obtained in this paper for quiescent and star-forming galaxies respectively, for the LEGA-C sample (open) and combined LEGA-C and SDSS sample (solid). Black symbols show results from other studies of quiescent galaxies.}
    \label{fig:ML_lit}
\end{figure}

We show a comparison with previous measurements of the redshift evolution in $M_{\mathrm{dyn}}/L$ of quiescent galaxies in Fig.~\ref{fig:ML_lit}, where colored markers represent results obtained with the LEGA-C data and black symbols indicate different studies. Our result for the quiescent sample is consistent with the evolution of field galaxies in the rest-frame $B$-band measured by \citet{Treu2005} and \citet{Saglia2010,Saglia2016}, and slightly steeper than the bias-corrected measurement by \citet{vdWel2005}.

Other studies \citep[e.g.,][]{vdSande2014,Beifiori2017} deviate more significantly (typically $2-3\sigma$), 
which can largely be attributed to differences in the sample selection. Our selection generally differs from previous works in (i) the diversity of environment probed, with many studies focusing on galaxy clusters alone, or (ii) the mass range considered, as many studies have been limited to more massive galaxies.

\citet{VanDokkum2007} and \citet{Saglia2010} have shown that the redshift evolution in $M_{\mathrm{dyn}}/L$ differs for cluster and field galaxies. 
If we restrict our fit to only those LEGA-C galaxies which are classified as being cluster members \citep{Darvish2017}, we also find a marginally shallower evolution of $\Delta\log(M_{\mathrm{dyn}}/L_g)\propto(-0.83\pm0.18)z$ as compared to the full LEGA-C sample. 

Moreover, \citet{vdWel2005} and others \citep[e.g.,][]{Holden2010,Joergensen2013} find evidence for a mass-dependent evolution of $M_{\mathrm{dyn}}/L$, with low-mass galaxies evolving more rapidly than high-mass galaxies. We would therefore expect to find a steeper evolution for our sample ($\log(M_*/M_\odot) >10.5$) as compared with previous studies that typically select galaxies of $\log( M_*/M_\odot) \gtrsim 11$. We indeed find a mass dependence within our sample: if we fit only LEGA-C galaxies in the mass range $10.5<\log(M_*/M_\odot)<10.8$ or $\log(M_*/M_\odot )>11.2$, we find $\Delta\log(M_{\mathrm{dyn}}/L_g)\propto (-1.1\pm0.2)z$ and $\Delta\log(M_{\mathrm{dyn}}/L_g)\propto (-0.73\pm0.11)z $ respectively. 

Lastly, we note that the above measurements neglect the role of progenitor bias \citep{VanDokkum2001}: less massive galaxies tend to assemble and quench later than high-mass galaxies, such that galaxies of a fixed stellar mass at $z\sim0$ will be younger than those at $z\sim0.8$, and therefore also have a lower $M_{\mathrm{dyn}}/L_g$. However, a full treatment of this effect on the FP is beyond the scope of this work.

\subsection{Star-forming galaxies}\label{sec:zevo_LFP_SF}

The evolution of the star-forming population is significantly shallower than that of the quiescent population. Although the specific star formation rate (sSFR) decreases sharply toward $z\sim0$ \citep{Madau2014}, and $M_{\mathrm{dyn}}/L_g$ thus strongly increases, any low level of star formation will 
reduce the net increase in $M_{\mathrm{dyn}}/L_g$. Moreover, progenitor bias plays a significant role: while young galaxies enter the massive star-forming population toward low redshift, many of the older galaxies become quiescent. The net effect is therefore a shallower observed evolution in $M_{\mathrm{dyn}}/L_g$.

\section{Evolution in $M_{\mathrm{dyn}}/M_*$} \label{sec:MM_results}

We obtain the mass FP by replacing the surface brightness ($I_{\mathrm{e,g}}$) by the stellar mass surface density ($\Sigma_*=M_*/(2\pi R_{\mathrm{e}}^2)$):
\begin{equation}
     \log R_\mathrm{e}=\alpha\,\log\sigma+\beta\,\log\Sigma_*+\gamma,
     \label{eq:mfp}
\end{equation}
where $\alpha$ and $\beta$ describe the tilt, and $\gamma$ is the zero point. 
Following the approach of Section~\ref{sec:ML_results}, we adopt a fixed tilt of $\alpha=1.629$ and $\beta=-0.84$ \citep{HydeBenardi2009}. We again fit the zero point of the SDSS sample ($\gamma_0$) for the star-forming and quiescent population separately, and calculate the residual of the FP in $M_{\mathrm{dyn}}/M_*$ for the LEGA-C galaxies:

\begin{equation}
    \Delta\log(M_{\mathrm{dyn}}/M_*)\approx-\Delta\log\Sigma_*=\left(\Delta_{\mathrm{MFP}}-\gamma_0\right)/\beta,
    \label{eq:MM}
\end{equation}
where 
\begin{equation}
    \Delta_{\mathrm{MFP}}=\log R_\mathrm{e}-\alpha\,\log\sigma-\beta\,\log\Sigma_*.
\end{equation}

\begin{figure*}
    \centering
    \includegraphics[width=0.9\linewidth]{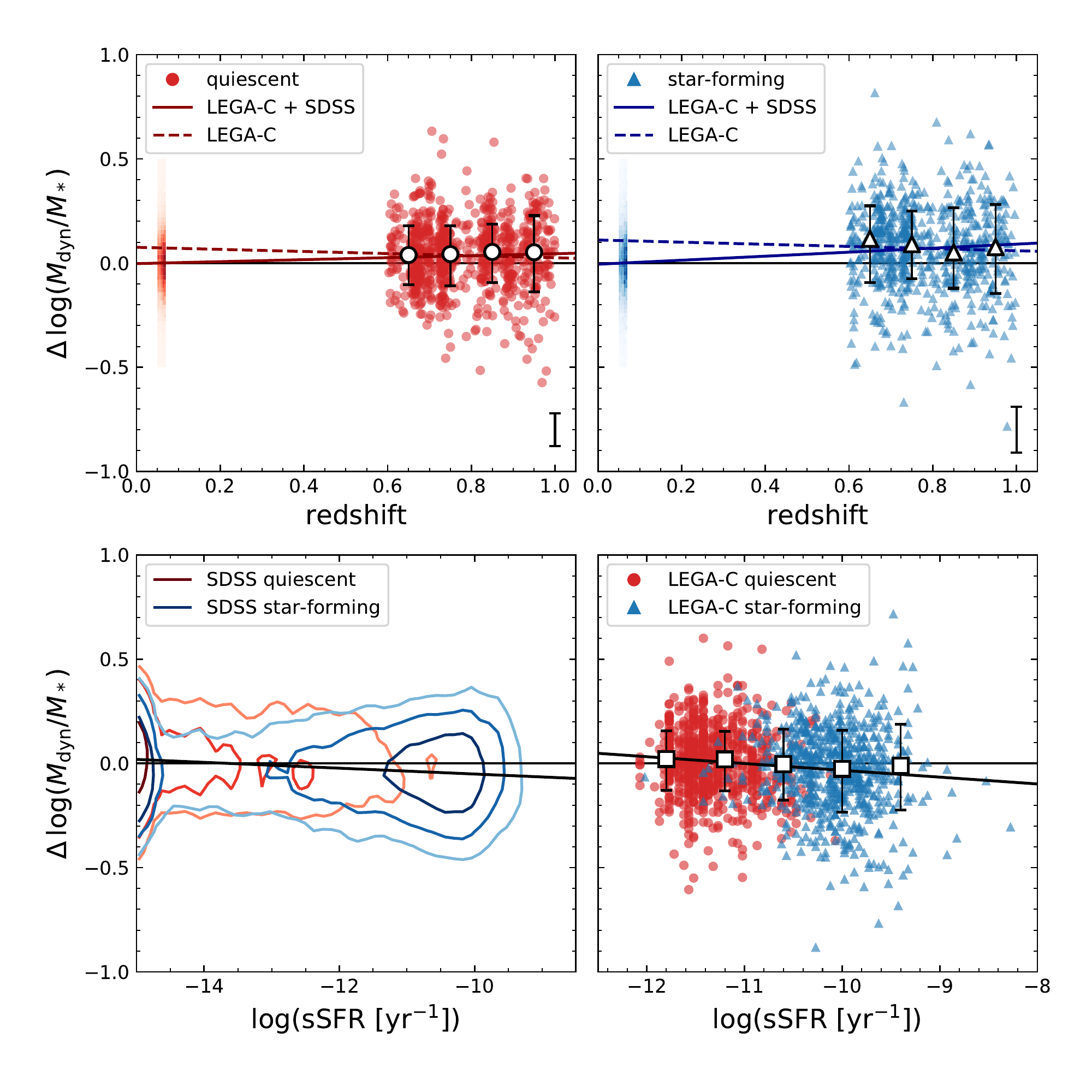}
    \caption{Evolution of the residuals from the mass FP of massive galaxies with redshift (top panels) and sSFR (bottom panels). Red and blue markers indicate the quiescent and star-forming LEGA-C galaxies respectively, with medians and 16\textsuperscript{th} and 84\textsuperscript{th} percentiles shown in black. The SDSS sample is represented by 2D histograms or contours that enclose $50\%$, $80\%$ and $95\%$ of each subsample (smoothed with a Gaussian filter of $\mathrm{FWHM=0.1\,dex}$). Linear fits to the LEGA-C data (dashed lines) and combined LEGA-C and SDSS sample (solid lines) show that the redshift evolution in $\Delta\log(M_{\mathrm{dyn}}/M_*)$ is, at most, weak (see Table~\ref{tab:zevo}). Combined with the very weak correlation between $\Delta\log(M_{\mathrm{dyn}}/M_*)$ and the sSFR (solid lines; bottom panels), this implies that massive star-forming and quiescent galaxies lie on the same mass FP across $0<z<1$.}
    \label{fig:MM_evo}
\end{figure*}

In Fig.~\ref{fig:MM_evo} we show $\Delta\log(M_{\mathrm{dyn}}/M_*)$ as a function of redshift for the star-forming (blue) and quiescent (red) LEGA-C and SDSS galaxies. 
As in Section~\ref{sec:ML_results}, we perform a linear fit to the two populations separately, using the LEGA-C data only (dashed lines) and the combined LEGA-C and SDSS data (solid lines). The results are presented in Table~\ref{tab:zevo}. 

For the quiescent galaxies, the two slopes are consistent within $1.6\sigma$, and agree well with the lack of evolution found by \citet{Bezanson2013} for a sample of $\sim100$ high-redshift quiescent galaxies. Our result demonstrates that the mass FP of the star-forming population also does not undergo a strong evolution. 

Furthermore, we demonstrate that this result is not sensitive to the adopted definition of quiescence. The bottom panels of Fig.~\ref{fig:MM_evo} show the dependence of $\Delta\log(M_{\mathrm{dyn}}/M_*)$ on the sSFR obtained from the SED fitting. There is only a weak correlation for both the SDSS and LEGA-C galaxies, as evidenced by linear fits to the data (black solid lines), with galaxies of high sSFR  being on average slightly more baryon-dominated within $1\,R_\mathrm{e}$: $\mathrm{d}\log(M_{\mathrm{dyn}}/M_*)/\mathrm{d}\log(\rm sSFR) = -0.014\pm0.0005$ and $\mathrm{d}\log(M_{\mathrm{dyn}}/M_*)/\mathrm{d}\log(\rm sSFR) = -0.033\pm0.007$ for the SDSS and LEGA-C samples respectively. 

The LEGA-C data alone suggest that all galaxies lie on the same mass FP, irrespective of star formation activity and redshift. However, both \citet{Schechter2014} and \citet{Zahid2016} find a weak redshift evolution in the zero point of the mass FP of early-type galaxies, such that $\Delta\log(M_{\mathrm{dyn}}/M_*)$ increases slightly with redshift. 
We find a similar weak but significant evolution in $M_{\mathrm{dyn}}/M_*$ with redshift for our combined LEGA-C and SDSS data, particularly so for the star-forming galaxies, raising the question of whether the observed evolution to $z\sim0$ is due to structural evolution, or caused by systematic uncertainties.

\begin{deluxetable}{lcc}
\tablenum{1}
\tablecaption{Best-fit evolution in $M_{\mathrm{dyn}}/L_g$ and $M_{\mathrm{dyn}}/M_*$}
\tablewidth{0pt}
\tablehead{
\colhead{Sample} & \colhead{$\mathrm{d}\log(M_{\mathrm{dyn}}/L_g)/\mathrm{d}z$} & \colhead{$\mathrm{d}\log(M_{\mathrm{dyn}}/M_*)/\mathrm{d}z$} 
}
%\decimalcolnumbers
\startdata
$0.6<z<1.0$ Q & $-0.86\pm0.07$ & $-0.05 \pm 0.06$  \\
$0.6<z<1.0$ SF & $-0.54\pm0.11$ & $-0.05 \pm 0.08$ \\
$0.0<z<1.0$ Q & $-0.728\pm0.011$ & $0.048 \pm 0.009$ \\
$0.0<z<1.0$ SF & $-0.604\pm0.016$ & $0.097 \pm 0.011$ \\
\enddata
\tablecomments{Samples correspond to either the LEGA-C data ($0.6<z<1.0$) or combined SDSS and LEGA-C data ($0.0<z<1.0$) for the quiescent (Q) and star-forming (SF) populations.}
\label{tab:zevo}
\end{deluxetable}

In Fig.~\ref{fig:ML_evo} we showed that the evolution of the luminosity FP is broadly consistent between the two data sets, suggesting that any systematic effects on the velocity dispersion, size, or luminosity are small. However, the stellar mass is an additional possible source of systematic error.
Although we have mitigated potential biases between the SDSS and LEGA-C data by using the same models and software for the SED modeling of all galaxies, we caution that some differences remain, particularly in the photometry used. For instance, the aperture sizes differ systematically, the SED is sampled differently in wavelength space, and there may be systematic uncertainties in the calibration of the photometry. 
Overall this can lead to a systematic uncertainty of at least $0.05\,$dex between the SDSS and LEGA-C mass estimates: for example, we find lower stellar masses for our SDSS sample if we use the MPA-JHU catalog \citep{Brinchmann2004}, with a median offset of $-0.05\,$dex compared to the masses from \citet{Chang2015}. This would shift the SDSS data upward in Fig.~\ref{fig:MM_evo}, in closer agreement with the LEGA-C data. 
We therefore conclude that the observed weak evolution in the mean value of $\Delta\log(M_{\mathrm{dyn}}/M_*)$ is likely not significant, and caution against interpreting this as evidence for, e.g., evolution in the dark matter fraction or the initial mass function.

\begin{figure*}
    \centering
    \includegraphics[width=0.9\linewidth]{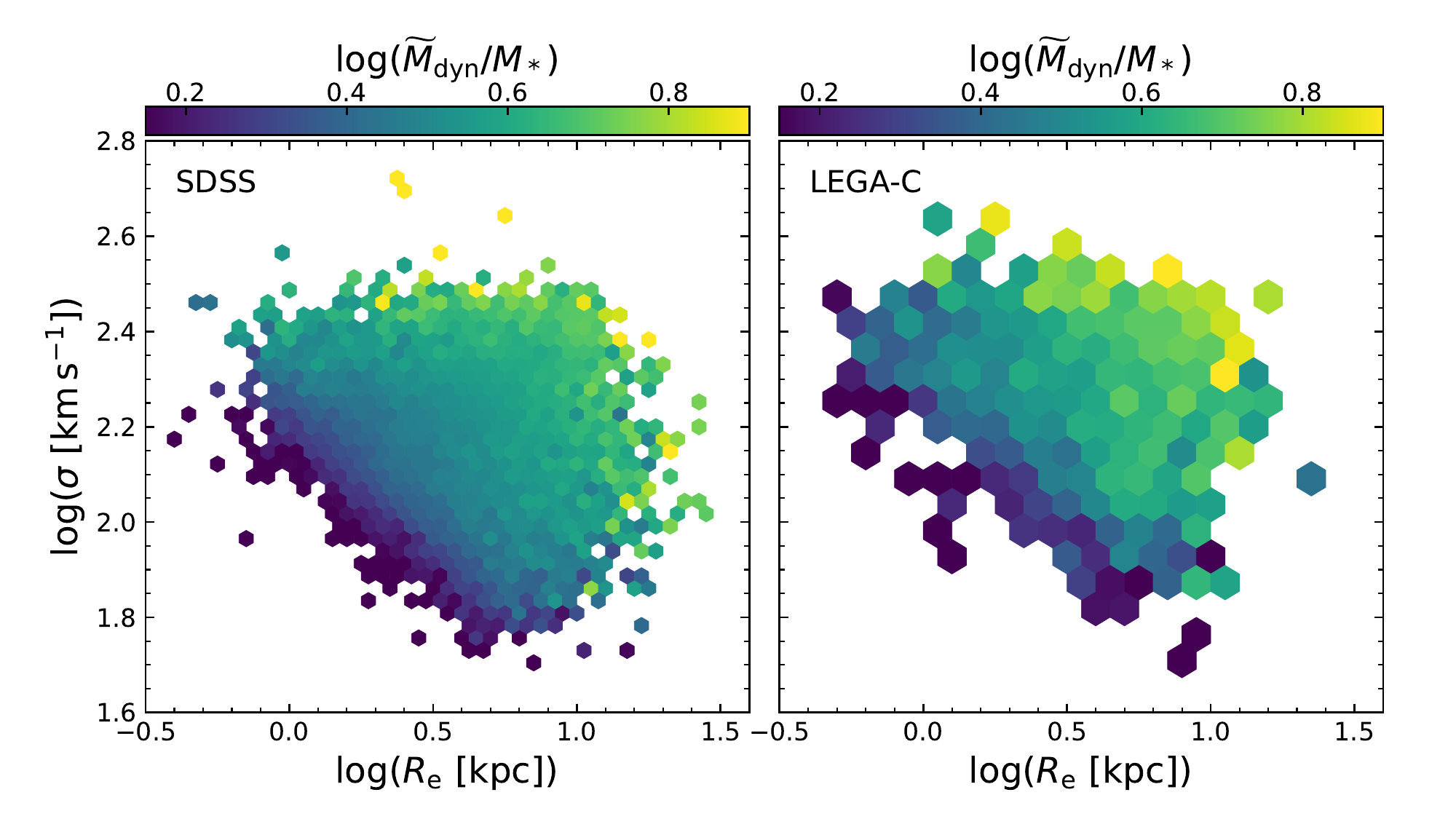}
    \caption{Variation in $M_{\mathrm{dyn}}/M_*$ along the mass FP. Panels show a near face-on projection of the mass FP, color-coded by the mean value of $\log(\widetilde{M}_{\mathrm{dyn}}/M_*)$ in each bin. Although Fig.~\ref{fig:MM_evo} shows no evolution in the mean value of $\Delta\log(M_{\mathrm{dyn}}/M_*)$ with either redshift or sSFR, it is possible for individual galaxies grow with time, and thus undergo a change in $M_{\mathrm{dyn}}/M_*$.} 
    \label{fig:faceon}
\end{figure*}

Systematics can also explain the discrepancy between our results and those by \citet{Bezanson2015}, who found that the mass FP changes by $\Delta\log(M_{\mathrm{dyn}}/M_*)\sim0.2-0.3\,$dex between $0\lesssim z\lesssim0.7$. However, the SED modeling differs significantly for their low-redshift and high-redshift data, resulting in a systematic offset: when using the same methods, i.e. stellar masses from the MPA-JHU catalog for the SDSS and masses estimated with FAST \citep{Kriek2009} for LEGA-C, we also find that $\mathrm{d}\log(M_{\mathrm{dyn}}/M_*)/\mathrm{d}z\approx0.3\,$dex.

Finally, we emphasize that although the residual from the FP in $M_{\mathrm{dyn}}/M_*$ is approximately constant across $0<z<1$ (Fig.~\ref{fig:MM_evo}), there is significant and systematic variation in $M_{\mathrm{dyn}}/M_*$ within the galaxy population itself. Fig.~\ref{fig:faceon} shows a near face-on projection of the mass FP color-coded by the mean value of $\log(\widetilde{M}_{\mathrm{dyn}}/M_*)$ in bins of $\log R_\mathrm{e}$ and $\log\sigma$, where $\widetilde{M}_{\mathrm{dyn}}$ is calculated following \citet{Cappellari2006}:
\begin{equation}
    \widetilde{M}_{\mathrm{dyn}}=\frac{\beta(n)R_\mathrm{e}\sigma^2}{G},
\end{equation}
with $\beta(n)=8.87-0.831n+0.0241n^2$, where $n$ is the S\'ersic index and $G$ the gravitational constant.  
While the zero point of the mass FP itself remains constant, individual galaxies may change in size and velocity dispersion with time, thus moving along the FP, and vary in $M_{\mathrm{dyn}}/M_*$.

\section{Discussion and Conclusions}

In this Letter, we have measured the redshift evolution of the luminosity and mass FP of massive ($\log(M_*/M_\odot)\geq10.5$) galaxies out to $z\sim1$.
Whereas previous studies suffered from significant selection bias, our sample of 1419 galaxies from the LEGA-C survey is highly homogeneous and representative of the massive galaxy population at $z\sim0.8$ \citep[][de Graaff et al. in prep.]{vdWel2016}. 
We find that the star-forming and quiescent populations follow a steep evolution in $M_{\mathrm{dyn}}/L_g$, yet, their evolution in $M_{\mathrm{dyn}}/M_*$ is remarkably weak: all massive galaxies lie on the same mass FP across $0<z<1$.

The stability of the mass FP implies that the evolution in the luminosity FP, and thus in $M_{\mathrm{dyn}}/L$, is due to a combination of progenitor bias and evolution in the stellar populations alone: $\Delta\log(M_{\mathrm{dyn}}/L)=\Delta\log(M_*/L)$.

There is some room for evolution of the mass FP with redshift, however, if we assume that the weak evolution in Fig.~\ref{fig:MM_evo} is physical, and not caused by systematic uncertainties. In this case, the weak dependence of the residuals from the FP on the sSFR and the different values of $\mathrm{d} \log(M_{\mathrm{dyn}}/M_*)/\mathrm{d}z$ for the star-forming and quiescent populations reflect structural differences.

In contrast, theoretical predictions \citep[e.g.,][]{Hilz2013} and observations \citep[e.g.,][]{vdSande2013,Wuyts2016,Genzel2017} show that---within the effective radius---galaxies become more baryon-dominated at high redshift, whereas the best-fit evolution of our combined LEGA-C and SDSS data suggests the opposite. We emphasize that systematic observational uncertainties likely contribute to the observed offset between the SDSS and data at higher redshift. Moreover, we note that we have not accounted for baryonic mass in the form of gas, which may become increasingly important toward high redshift. We have also not included the effect of color gradients, which may lead to an underestimation of $M_\mathrm{dyn}/M_*$, since mass-weighted sizes can be substantially smaller than the luminosity-weighted sizes used here \citep[e.g.,][]{Szomoru2013,Chan2016}.

The lack of evolution of the mass FP implies that the coupling of morphological and dynamical properties extends over a wide range in time, imposing strong constraints on the possible evolutionary pathways of galaxies. Quiescent galaxies for example, which have been shown to undergo significant size growth between $0<z<1$ \citep{vdWel2014}, must evolve dynamically such as to remain on the mass FP (Fig.~\ref{fig:faceon}). 

Moreover, we find that the star-forming progenitors lie on the same scaling relation as their massive, quiescent descendants at low redshift. The mass FP therefore offers a tool to study the structural and kinematic evolution of galaxies with minimal impact from progenitor bias, by statistically tracking their trajectories along the plane. 

Whether the mass FP can be used in a similar fashion at $z>1$ or at lower mass, will require a larger number of stellar kinematic measurements at high redshift. Future studies will help to understand how galaxies settle onto the scaling relation, and whether galaxies become more baryon-dominated at high redshift.

\acknowledgments

Based on observations made with ESO Telescopes at the La Silla Paranal Observatory under program ID 194-A.2005 (The LEGA-C Public Spectroscopy Survey). This project has received funding from the European Research Council (ERC) under the European Union’s Horizon 2020 research and innovation program (grant agreement No. 683184). We gratefully acknowledge the NWO Spinoza grant. CP is supported by the Canadian Space Agency under a contract with NRC Herzberg Astronomy and Astrophysics. JvdS acknowledges support of an Australian Research Council Discovery Early Career Research Award (project number DE200100461) funded by the Australian Government. PFW acknowledges the support of the fellowship from the East Asian Core Observatories Association.

%TC:ignore
% \quickwordcount{main}
% \quickcharcount{main}
% \detailtexcount{main}
%TC:endignore

\bibliography{legac}{}
\bibliographystyle{aasjournal}



\end{document}